\documentclass[twocolumn,amsmath,amssymb,osajnl]{revtex4}
\usepackage{epsfig}

\begin{document}

\title{Reduced micro-deformation attenuation in\\ large-mode area photonic crystal fibers for visible applications}

\author{Martin D. Nielsen}
\affiliation{Crystal Fibre A/S, Blokken 84, DK-3460 Birker\o d, Denmark\\Research Center COM, Technical University of Denmark, DK-2800 Kongens Lyngby, Denmark}

\author{Niels Asger Mortensen}
\affiliation{Crystal Fibre A/S, Blokken 84, DK-3460 Birker\o d, Denmark}

\author{Jacob Riis Folkenberg}
\affiliation{Crystal Fibre A/S, Blokken 84, DK-3460 Birker\o d, Denmark}

\date{\today}

\begin{abstract}
We consider large-mode area photonic crystal fibers for visible applications where micro-deformation induced attenuation becomes a potential problem when the effective area $A_{\rm eff}$ is sufficiently large compared to $\lambda^2$. We argue how a slight increase in fiber diameter $D$ can be used in screening the high-frequency components of the micro-deformation spectrum mechanically and we confirm this experimentally for both $15\,{\rm \mu m}$ and $20\,{\rm \mu m}$ core fibers. For typical bending-radii ($R\sim 16\,{\rm cm}$) the operating band-width increases by $\sim 3-400\,{\rm nm}$ to the low-wavelength side.
\end{abstract}


\maketitle

\begin{figure}[b!]
\begin{center}
\epsfig{file=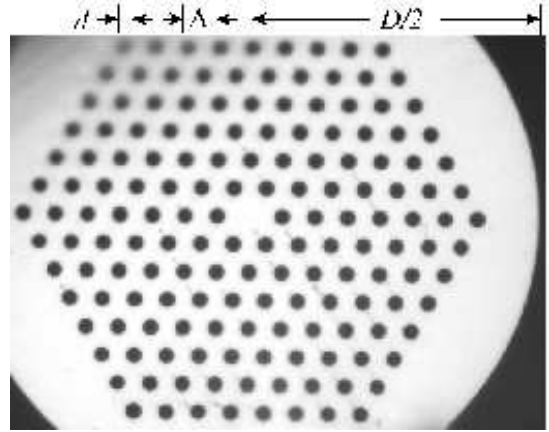, width=0.4\textwidth,clip}
\end{center}
\caption{Optical micrograph of cross-section of a PCF with a triangular arrangement of air holes (dark regions) and a $15\,{\rm \mu m}$ silica core. The outer diameter is $D\simeq 175\,{\rm \mu m}$ and the air holes are of diameter $d/\Lambda\simeq 0.44$ with a pitch $\Lambda\simeq 10\,{\rm \mu m}$. }
\label{fig1}
\end{figure}

In all-silica photonic crystal fibers (PCFs) guidance of light is provided by an arrangement of air-holes running along the full length of the fiber. Typically, the air-holes of diameter $d$ are arranged in a triangular lattice with a pitch $\Lambda$ of the same length scale as the free-space wavelength $\lambda$. The core can either be a solid silica-core (see Fig.~\ref{fig1}) with total-internal reflection guidance\cite{knight1996} or an air-core with photonic-bandgap guidance.\cite{cregan1999} For recent reviews we refer to Ref.~\onlinecite{russell2003} and references therein. 

Though the air-core PCFs probably have the most extraordinary guidance mechanism, the silica-core PCFs also have remarkable properties such as their endlessly single-mode nature\cite{birks1997} leading to in principle unlimited large effective areas.\cite{knight1998el} These properties are often greatly desired for high-power delivery and this makes PCFs obvious candidates for many of these applications. 
For the single-mode PCFs there is of course a tail of the coin common to the physics in standard-fiber technology; as the effective area $A_{\rm eff}$ is increased the mode becomes increasingly susceptible to longitudinal fiber variations, micro-deformations, and macro-bending.\cite{mortensen2003b} However, despite this the PCF technology has a clear advantage because of the endlessly single-mode properties and the relative ease by which small effective core-cladding index steps $\Delta n_{\rm eff}$ can be realized through control of $d$ and $\Lambda$. In standard fibers the latter is difficult since extremely small well-controlled doping-levels are required to achieve $\Delta n_{\rm eff} \ll 10^{-3}$.

Even though the general physics is that $\Delta n_{\rm eff}$ decreases if $A_{\rm eff}$ is increased there is still room for improvement by optimal choice of the micro-structured cladding and the core-shape and size. Recently we demonstrated an improved large-mode area (LMA) design with a three-rod core\cite{mortensen2003a} instead of the usual one-rod core.\cite{knight1996} Compared to the one-rod design this new LMA-PCF offers a larger $\Delta n_{\rm eff}$ for the same $A_{\rm eff}$ or vice versa in the $\lambda \ll \Lambda$ limit.

The general view has been that the operation of LMA-PCFs is limited by macro-bending loss,\cite{knight1998el,baggett2001,sorensen2001} but pushing the technology to still larger effective areas limitations can be set by {\it e.g} micro-bending deformations as well.\cite{mortensen2003b} Micro-bending deformations may be caused by external perturbations as studied recently,\cite{nielsen2003} but even when there is no external perturbations there may still be residual micro-deformations caused by frozen-in mechanical stress in {\it e.g.} the coating material. In the situation where performance is not limited by macro-bending it is obvious to look into properties of the screening of micro-bending deformations. Recently, results for the attenuation of a PCF with a $15\,{\rm \mu m}$ core diameter and an outer diameter of $D=125\,{\rm \mu m}$ were reported.\cite{mortensen2003b} In the visible regime the performance of this PCF was clearly found to suffer from micro-deformation induced attenuation. 
In this Letter we suggest a simple way of screening the effect of micro-deformations and demonstrate a considerable reduction of the loss-level in LMA PCFs for visible applications. We emphasize that the actual loss-level arises from competing effects\cite{mortensen2003b} and once the effect of micro-bending has been suppressed below the macro-bending induced attenuation-level the operation of the fiber may of course still be limited by macro-bending attenuation.

The Fourier spectrum of naturally occurring micro-bending deformations will typically be quite broad-band although sharp distinct features may also be realized {\it e.g.} for periodic deformations.\cite{nielsen2003} In either case the high-frequency components of the externally applied deformation spectrum do not affect the optical waveguide, because of the stiffness of the fiber. In other words it is practically impossible to deform the fiber on a length scale comparable to or shorter than the fiber diameter. In terms of the effective index step micro-deformations will be screened if $\Delta n_{\rm eff} > \Delta n_{\rm micro}$. For a simple fiber-model it can be shown that \cite{bjarklev1989}

\begin{equation}\label{micro}
\Delta n_{\rm micro}=(\lambda/\pi D)(E_c/\pi E_f)^{1/4}
\end{equation}
where $D$ is the fiber diameter, $E_c$ is the Young's modulus of the coating, and $E_f$ is the effective Young's modulus of the air-silica composite material. In this work we focus on the $D$-dependence, but changing the screening by use of coating with more advanced mechanical properties is an alternative direction.

From Eq.~(\ref{micro}) it is obvious that the $\Delta n_{\rm micro}\propto 1/D$ property can be used as a simple way to screen micro-deformations. In the optical fiber-community it is a long-standing tradition and standard to use $D\simeq 125\,{\rm \mu m}$, but as we will demonstrate there is a clear advantage of increasing the diameter of LMA-PCFs (for {\it e.g.} visible applications) similarly to what is often done for standard technology specialty fibers, see {\it e.g.} Ref.~\onlinecite{fermann1998}.

Using numbers for silica and typical polymer coatings Eq.~(\ref{micro}) suggest that $\Delta n_{\rm micro} \sim 0.1 \times \lambda/D$. For typical fiber diameters and wavelengths this gives numbers which in order of magnitude only differ slightly from typical mode-spacings.\cite{nielsen2003} For the PCFs we are going to compare we have two versions with $D=125\,{\rm \mu m}$ and $175\,{\rm \mu m}$, respectively, but with otherwise similar dielectric cross-sections in terms of core-size, pitch, and air-hole diameter. Though the change in diameter may seem like a modest change it has an important effect on $\Delta n_{\rm micro}$ which is lowered by $\sim30\,\%$ and for the PCFs studied in this work we shall see that this change is enough to bring the PCF from the $\Delta n_{\rm eff}< \Delta n_{\rm micro}$ regime to the $\Delta n_{\rm eff}> \Delta n_{\rm micro}$ regime. For PCFs $\Delta n_{\rm eff}$ decreases when the wavelength decreases (the opposite to the case in standard fibers) and thus micro-bending loss will increase with decreasing wavelength.

\begin{figure}[t!]
\begin{center}
\epsfig{file=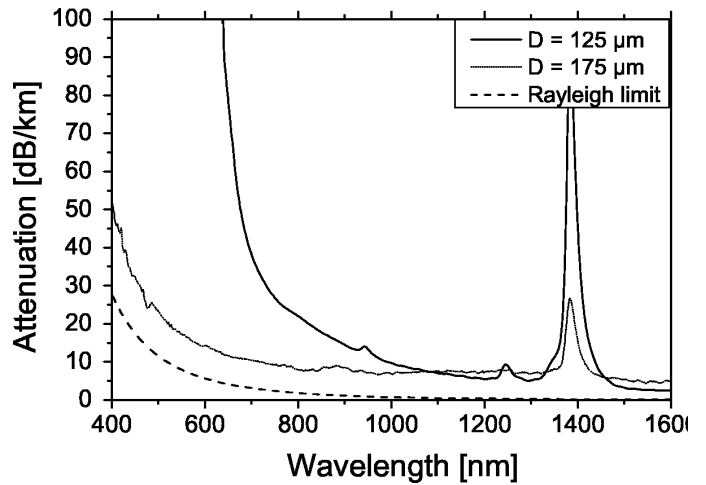, width=0.5\textwidth,clip}
\end{center}
\caption{Spectral attenuation for two PCFs both with a $15\,{\rm \mu m}$ core diameter characterized on a spool of $8\,{\rm cm}$ radius. The solid line shows results for an outer diameters of $D=125\,{\rm \mu m}$, the dotted line is for $D=175\,{\rm \mu m}$, and the dashed line indicates the fundamental attenuation limit set by Rayleigh scattering.}
\label{fig2}
\end{figure}

In Fig.~\ref{fig2} we show the experimentally observed spectral attenuation for two endlessly single-mode PCFs with a core diameter $2\Lambda-d$ of $15\,{\rm \mu m}$ and with an air-hole diameter $d/\Lambda\sim 0.44$, {\it i.e.} close to the theoretical endlessly single-mode limit.\cite{kuhlmey2002} The fibers are drawn from similar preforms fabricated by the stack-and-pull method,\cite{knight1996} but different thicknesses of the over-cladding were used for the two drawings. This results in two fibers with similar micro-structures, but with the different values of $D$ as mentioned above. The fibers have 5 and 7 rings of air holes, respectively, but since $\lambda\ll\Lambda$ this has no importance in terms of leakage loss\cite{white2001} and thus the different thicknesses of over-cladding does not influence the optical properties of guided modes. The two fibers were both found to be single-mode (in both the visible and the near-infrared regimes) and they have similar mode-field diameters ($\sim 12\,{\rm \mu m}$) and otherwise only differ in their mechanical properties. For each of the fibers the attenuation was characterized with the cut-back technique using a fiber length of $> 100\,{m}$ on a spool of radius $8\,{\rm cm}$, a white-light source, and an optical spectrum analyzer. Comparing the two curves in Fig.~\ref{fig2} the effect of a larger outer diameter $D$ is seen to be quite dramatic and for the PCF with the largest diameter the spectral dependence has a typical Rayleigh dependence though there is a wavelength independent offset compared to the ultimate Rayleigh limit which is indicated by the dashed line. This indicates the absence of both macro-bending and micro-deformation induced attenuation. The slightly higher loss of the $D=175\,{\rm \mu m}$ for $\lambda \gtrsim 1100\,{\rm nm}$ originates in contamination.

\begin{figure}[t!]
\begin{center}
\epsfig{file=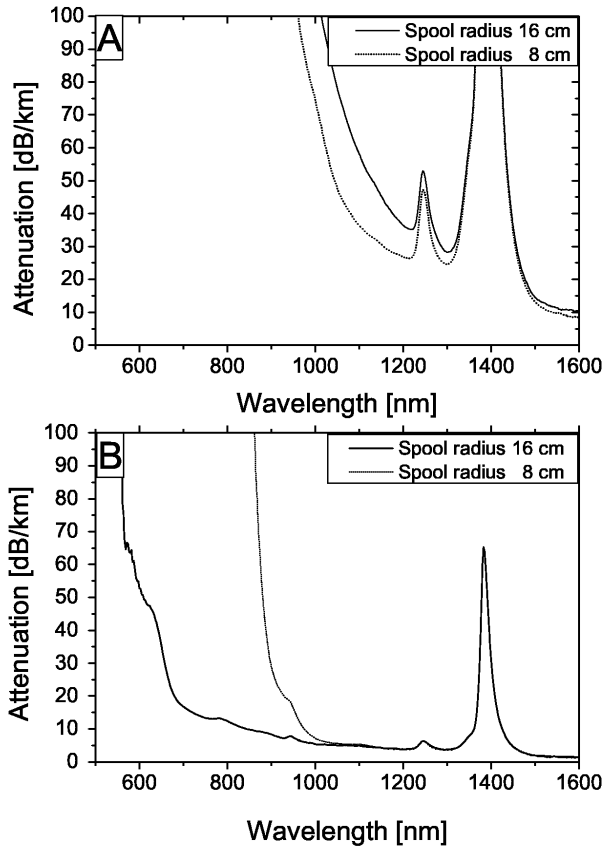, width=0.5\textwidth,clip}
\end{center}
\caption{Spectral attenuation for different macro-bending for two PCFs with a $20\,{\rm \mu m}$ core diameter. In panels A and B the outer diameter of the PCF is $D=125\,{\rm  \mu m}$ and $230\,{\rm \mu m}$, respectively.}
\label{fig3}
\end{figure}

In Fig.~\ref{fig3} we consider two PCFs with a $20\,{\rm \mu m}$ core diameter and $d/\Lambda\sim 0.44$ and compare two versions; one with $D=125\,{\rm  \mu m}$ where the effect of micro-deformations is pronounced and one with $230\,{\rm \mu m}$ where the effect of the deformation is screened. The fibers have 3 and 7 rings of air-holes, respectively, but we emphasize that this difference has no importance in terms of leakage loss since $\lambda\ll\Lambda$.\cite{white2001} Again the version with a larger outer diameter has a significant lower attenuation-level (compare panels A and B). By changing the macro-bending radius (see panel B) we actually find evidence that the effect of micro-deformations is fully screened even $200\,{\rm nm}$ below the O-H attenuation peak at $\lambda \sim 1.24\,{\rm \mu m}$ and that macro-bending limitations become more apparent. For the PCF with the smallest outer diameter the attenuation-level caused by micro-deformations is too high to make macro-bending a limiting effect (see panel A) whereas we for the large-diameter version observe a clear bend-edge which as expected shifts toward longer wavelengths when decreasing the bend-radius. In panel A the attenuation is actually seen to be higher on the $16\,{\rm cm}$ spool further indicating that micro-deformation is the limiting factor. The difference is caused by different spooling-conditions, {\it e.g.} the tension or the physical surface-topography of the spool.

In conclusion we have shown how micro-deformation induced loss in LMA-PCFs may be suppressed by a slight increase in fiber diameter $D$ compared to the standard of $D\sim 125\,{\rm \mu m}$. The suppression can be understood as a simple mechanical screening of the high-frequency components of the micro-deformation spectrum. For a PCF with a $20\,{\rm \mu m}$ core we have demonstrated a significant reduction of the loss-level by going from $D\simeq 125\,{\rm \mu m}$ to $D\simeq 230\,{\rm \mu m}$. For a bend-radius $R\simeq 16\,{\rm cm}$ this improved PCF guides light with a loss-level not exceeding $\sim 20\,{\rm dB/km}$ down to $\lambda\sim 650\,{\rm nm}$ where macro-bending attenuations sets in. We believe the present findings have important implications for the perspectives of utilizing LMA-PCFs for high-power delivery in the visible regime and for use of the PCF technology in fiber-laser applications.

\vspace{5mm}

M.~D. Nielsen acknowledges financial support by the Danish Academy of Technical Sciences. M.~D. Nielsen's e-mail address is mdn@crystal-fibre.com.


\end{document}